\documentclass[twocolumn,aps,superscriptaddress,showpacs]{revtex4}

\usepackage{amssymb}
\usepackage{amsmath}
\usepackage{graphicx}
\usepackage[normalem]{ulem}
\usepackage[dvips]{color}

\begin{document}

\title{Shear viscosity of nuclear matter}
\author{Jun Xu}\email{xujun@sinap.ac.cn}
\affiliation{Department of Physics and Astronomy, Texas A$\&$M
University-Commerce, Commerce, TX 75429-3011, USA}
\affiliation{Shanghai Institute of Applied Physics, Chinese Academy
of Sciences, Shanghai 201800, China}

\date{\today}

\begin{abstract}

In this talk I report my recent study on the shear viscosity of
neutron-rich nuclear matter from a relaxation time approach. An
isospin- and momentum-dependent interaction is used in the study.
Effects of density, temperature, and isospin asymmetry of nuclear
matter on its shear viscosity have been discussed. Similar to the
symmetry energy, the symmetry shear viscosity is defined and its
density and temperature dependence are studied.

\end{abstract}


\maketitle

\section{Introduction}
\label{introduction}

One of the major problems in nuclear physics is to understand the
properties of nuclear matter under extreme conditions. This is
related to the basic knowledge of the in-medium nucleon-nucleon (NN)
interaction which in the present stage can still hardly be obtained
from the ab initio theory of the strong interaction, i.e., Quantum
chromodynamics. Our knowledge on the in-medium NN interaction today
is mainly developed along two lines. In the first line, one starts
from the bare NN interaction, which has been fitted very well from
NN scattering data, together with phenomenological three-body
interactions, so that the in-medium NN interaction and the
properties of nuclear matter can be obtained through many-body
theories. In the second line, the starting point is an effective
in-medium NN interaction or Lagrangian, with the parameters fitted
to the empirical nuclear matter properties obtained usually through
mean-field approximations.

Ten years ago, an isospin- and momentum-dependent mean-field
potential (hereafter 'MDI') was constructed to study the dynamics
(especially the isospin effects) in intermediate-energy heavy-ion
collisions together with an isospin-dependent
Boltzmann-Uehling-Uhlenbeck (IBUU) transport model~\cite{Das03}. In
addition to the good description of the empirical nuclear equation
of states, the momentum dependence of this mean-field potential
reproduces pretty good the optical potential extracted by Hama {\it
et al.} from elastic proton scattering data~\cite{Ham90}. The
studies using this interaction have constrained the nuclear symmetry
energy at both subsaturation and suprasaturation
densities~\cite{Che05,Li05,Xia09}. In addition to the dynamics of
heavy-ion collisions, the MDI model has also been used to the study
the thermodynamical properties of nuclear matter~\cite{Xu07,Xu08a}.
It was recently found that the isospin- and momentum-dependent
potential can be derived from an effective interaction with a
density-dependent two-body interaction and a Yukawa-type
finite-range interaction using Hartree-Fock calculation~\cite{Xu10}.
The MDI model thus serves as a useful effective in-medium
interaction.

In the past few years, the shear viscosity of the quark-gluon plasma
(QGP) formed in relativistic heavy-ion collisions has attracted
special attentions. From the study with a viscous hydrodynamical
model~\cite{Son11}, it was found that the strong-interacting QGP
behaves like a nearly ideal fluid, i.e., its specific shear
viscosity is only a little larger than the KSS
boundary~\cite{Kov05}. Up to now large efforts have been devoted to
study the shear viscosity of QGP~\cite{Pes05,Maj07,Xu08,Che10} and
hadron resonance gas~\cite{Mur04,Che07a,Dem09,Pal10a} formed in
relativistic heavy-ion collisions, while there are only a few
studies on the shear viscosity of nuclear matter formed in
intermediate-energy heavy-ion
collisions~\cite{Dan84,Shi03,Che07b,Pal10b,Li11}. Even few studies
are related to the isospin effects on the shear viscosity of nuclear
matter~\cite{Zha10}. In the present talk I will discuss my recent
study~\cite{Xu11} on the shear viscosity of nuclear matter using the
MDI model mentioned above from a relaxation time approach, which
gives an intuitive picture how the shear viscosity changes with the
density, temperature, and isospin asymmetry of nuclear matter.

\section{Shear viscosity from a relaxation time approach}
\label{viscosity}

The system concerned here is an isospin asymmetric nuclear matter
with uniform neutron and proton density $\rho_n$ and $\rho_p$,
respectively, and the nucleons are thermalized with temperature $T$.
The flow field $\vec{u}$ is static in the $z$ direction and its
magnitude is linear in the coordinate $x$, i.e., $u_z=cx$ and
$u_x=u_y=0$. In the rest frame nucleons move with the flow field and
follow Fermi-Dirac distribution $n^\star$ in the equilibrium state.
In the lab frame the equilibrium distribution is a simple boost by
the flow field compared with that in the rest frame, denoted as
$n^0$. Due to NN collisions, the real distribution may be slightly
away from the equilibrium distribution and is denoted as $n$, and
the deviation from the equilibrium distribution $\delta n=n^0-n$ is
much smaller than $n^0$.

The shear force between flow layers per unit area by definition can
be written as
\begin{eqnarray}\label{F}
\frac{F}{A} &=& \sum_\tau <(p_z-mu_z)\rho_\tau v_x>.
\end{eqnarray}
In the above, $\tau=n$ or $p$ denotes the isospin degree of freedom,
$\rho_\tau v_x$ is the number of nucleons moving between layers per
unit time per unit area, and $p_z-mu_z$ is the momentum transfer per
nucleon in the $z$ direction. The nucleon velocity in the $x$
direction $v_x$ can be further written as $v_x=p_x/m_\tau^\star$,
with $m_\tau^\star$ being the effective mass. Using the momentum
distribution $n_\tau=n_\tau^0+\delta n_\tau$ to calculate the
average and taking into account that the equilibrium momentum
distribution $n_\tau^0$ is even in $p_x$, Eq.~(\ref{F}) can be
further written as
\begin{equation}\label{shearF}
\frac{F}{A} = \sum_\tau d \int (p_z-mu_z)\frac{p_x}{m^\star_\tau}
\delta n_\tau \frac{d^3p}{(2\pi)^3},
\end{equation}
where $d=2$ is the spin degeneracy.

\begin{widetext}
In the following I will calculate $\delta n_\tau$ by linearizing the
isospin-dependent BUU equation as follows
\begin{eqnarray}\label{BUU}
&&\frac{\partial n_\tau(p_1)}{\partial t} + \vec{v} \cdot \nabla_r
n_\tau(p_1) - \nabla_r U_\tau \cdot \nabla_p n_\tau(p_1) = -
(d-\frac{1}{2})\int \frac{d^3p_2}{(2\pi)^3}
\frac{d^3p_1^\prime}{(2\pi)^3} \frac{d^3p_2^\prime}{(2\pi)^3}
|T_{\tau,\tau}|^2 \notag \\
&\times&
[n_\tau(p_1)n_\tau(p_2)(1-n_\tau(p_1^\prime))(1-n_\tau(p_2^\prime))-
n_\tau(p_1^\prime)n_\tau(p_2^\prime)(1-n_\tau(p_1))(1-n_\tau(p_2))] \notag\\
&\times& (2\pi)^3
\delta^{(3)}(\vec{p}_1+\vec{p}_2-\vec{p}_1^\prime-\vec{p}_2^\prime)-
d\int \frac{d^3p_2}{(2\pi)^3} \frac{d^3p_1^\prime}{(2\pi)^3}
\frac{d^3p_2^\prime}{(2\pi)^3}
|T_{\tau,-\tau}|^2 \notag \\
&\times&
[n_\tau(p_1)n_{-\tau}(p_2)(1-n_\tau(p_1^\prime))(1-n_{-\tau}(p_2^\prime))-
n_\tau(p_1^\prime)n_{-\tau}(p_2^\prime)(1-n_\tau(p_1))(1-n_{-\tau}(p_2))] \notag\\
&\times& (2\pi)^3
\delta^{(3)}(\vec{p}_1+\vec{p}_2-\vec{p}_1^\prime-\vec{p}_2^\prime).
\end{eqnarray}
In the above, $T$ is the transition matrix, the degeneracy $d-1/2$
takes the double counting of identical nucleon collisions into
consideration, and $1-n$ is from the Pauli blocking effect.
Replacing $n$ with $n^0$ in the first-order approximation, the
left-hand side can be expressed as
\begin{eqnarray}\label{LHS}
\frac{\partial n_\tau(p_1)}{\partial t} + \vec{v} \cdot \nabla_r
n_\tau(p_1) - \nabla_r U_\tau \cdot \nabla_p n_\tau(p_1) =
\left(-\frac{\partial u_z}{\partial x} \frac{p_z p_x}{p} \frac{d
n^0_\tau}{dp}\right)_{p=p_1}
\end{eqnarray}
by using the properties of $n^0$. Keeping only the $\delta n_\tau
(p_1)$ term, the right-hand side of Eq.~(\ref{BUU}) can be expressed
as $\delta n_\tau (p_1)/\tau_\tau(p_1)$, where $\tau_\tau(p_1)$ is
the relaxation time, i.e., the average time between two collisions
for a nucleon with isospin $\tau$ and momentum $p_1$, and it can be
written as
\begin{equation}\label{tauds}
\frac{1}{\tau_\tau(p_1)} = \frac{1}{\tau_\tau^{same}(p_1)} +
\frac{1}{\tau_\tau^{diff}(p_1)},
\end{equation}
where $\tau_\tau^{same(diff)}(p_1)$ is the average time for a
nucleon with momentum $p_1$ to collide with other nucleons of same
(different) isospin, and they can be calculated respectively from
\begin{eqnarray}
\frac{1}{\tau_\tau^{same}(p_1)} &=&
(d-\frac{1}{2})\frac{(2\pi)^2}{(2\pi)^3} \int p_2^2 dp_2
d\cos\theta_{12} d\cos\theta
\frac{d\sigma_{\tau,\tau}}{d\Omega}\left|\frac{\vec{p}_1}{m_\tau^\star(p_1)}-\frac{\vec{p}_2}{m_\tau^\star(p_2)}\right|
\notag\\
&\times&\left[n^0_{\tau}(p_2)-n^0_{\tau}(p_2)n^0_\tau(p_1^\prime)-n^0_{\tau}(p_2)n^0_{\tau}(p_2^\prime)+n^0_\tau(p_1^\prime)n^0_{\tau}(p_2^\prime)\right],\label{tausame}\\
\frac{1}{\tau_\tau^{diff}(p_1)} &=& d\frac{(2\pi)^2}{(2\pi)^3} \int
p_2^2 dp_2 d\cos\theta_{12} d\cos\theta
\frac{d\sigma_{\tau,-\tau}}{d\Omega}\left|\frac{\vec{p}_1}{m_\tau^\star(p_1)}-\frac{\vec{p}_2}{m_{-\tau}^\star(p_2)}\right|
\notag\\
&\times&\left[n^0_{-\tau}(p_2)-n^0_{-\tau}(p_2)n^0_\tau(p_1^\prime)-n^0_{-\tau}(p_2)n^0_{-\tau}(p_2^\prime)+n^0_\tau(p_1^\prime)n^0_{-\tau}(p_2^\prime)\right].\label{taudiff}
\end{eqnarray}
\end{widetext}
In the above $\theta_{12}$ is the angel between $\vec{p}_1$ and
$\vec{p}_2$, and $\theta$ is the scattering angel between the total
momentum and the relative momentum of the final state. In free space
the pp and np scattering cross sections are isotropic and they can
be respectively parameterized as~\cite{Cha90}
\begin{eqnarray}
&&\sigma_{pp(nn)} = 13.73 - 15.04/v + 8.76/v^2 +
68.67v^4,\label{sigma1}\\
&&\sigma_{np} = -70.67 - 18.18/v + 25.26/v^2 +
113.85v,\label{sigma2}
\end{eqnarray}
where the cross sections are in mb and $v$ is the velocity of the
projectile nucleon with respect to the fixed target nucleon. It is
worth to note that in the most probable collision energies the np
scattering cross section is about three times the pp scattering
cross section. In nuclear matter, the in-medium NN scattering cross
sections are modified by the in-medium effective mass in the form of
~\cite{Li05}
\begin{equation}
\sigma^{medium}_{NN} =
\sigma_{NN}\left(\frac{\mu_{NN}^\star}{\mu_{NN}}\right)^2,
\end{equation}
where $\mu_{NN}$ ($\mu_{NN}^\star$) is the free-space (in-medium)
reduced mass of colliding nucleons.

Once the relaxation time $\tau_\tau(p)$ is known, $\delta n_\tau(p)$
can be calculated from
\begin{equation}\label{deltan}
\delta n_\tau(p) = \tau_\tau(p) \frac{\partial u_z}{\partial x}
\frac{p_z p_x}{p} \frac{d n^0_\tau}{dp}.
\end{equation}
Using the definition $F/A = -\eta (\partial u_z/\partial x)$, the
shear viscosity can be calculated from Eqs.~(\ref{shearF}) and
(\ref{deltan}) in terms of the local momentum distribution
$n^\star_\tau$ as
\begin{eqnarray}\label{eta}
\eta = \sum_\tau -d \int \tau_\tau(p) \frac{p_z^2
p_x^2}{pm^\star_\tau} \frac{d n^\star_\tau}{dp}
\frac{d^3p}{(2\pi)^3},
\end{eqnarray}
by setting the magnitude of the velocity field to be infinitely
small. Note that from Eq.~(\ref{eta}) the shear viscosity is related
to the local momentum distribution near the Fermi surface.

\section{Results and discussions}
\label{results}

\begin{figure}[t]
\includegraphics[scale=0.8]{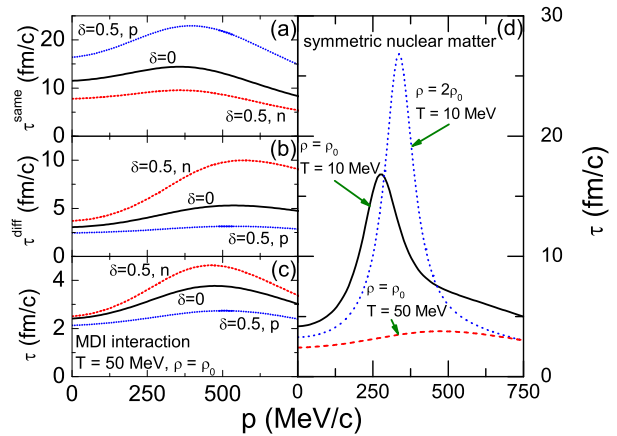}
\caption{(color online) Panel (a), (b), (c): Relaxation time for
neutrons and protons as a function of nucleon momentum in symmetric
($\delta=0$) and asymmetric ($\delta=0.5$) nuclear matter at
saturation density and temperature $T=50$ MeV; Panel (d): Relaxation
time as a function of nucleon momentum in symmetric nuclear matter
at different densities and temperatures.} \label{tau}
\end{figure}

Figure \ref{tau} displays the density, temperature, and isospin
dependence of the relaxation time. In neutron-rich nuclear matter,
$\tau_n^{diff}$ is larger while $\tau_n^{same}$ is smaller compared
to that in symmetric nuclear matter as a result of less frequent np
collisions and more frequent nn collisions. For the similar reason,
$\tau_p^{same}$ is larger while $\tau_p^{diff}$ is smaller compared
to that in symmetric nuclear matter. From Eq.~(\ref{tauds}), the
total relaxation time is determined by $\tau^{diff}$ which is always
smaller than $\tau^{same}$ due to the larger np cross section than
pp (nn) cross section in the most probably collision energies. Thus,
neutrons have a larger relaxation time than protons in neutron-rich
nuclear matter. It is seen in Panel (d) that the relaxation time
decreases with increasing temperature due to more frequent
collisions at higher temperatures. In addition, at lower
temperatures the relaxation time peaks around the Fermi momentum,
indicating a strong Pauli blocking effect for nucleons near the
Fermi surface.

\begin{figure}[t]
\includegraphics[scale=0.8]{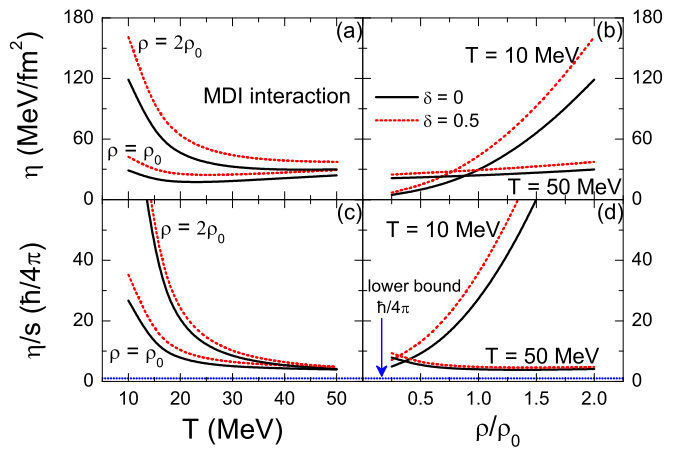}
\caption{(color online) Density and temperature dependence of the
shear viscosity ((a), (b)) and specific shear viscosity ((c), (d))
for symmetric ($\delta=0$) and asymmetric ($\delta=0.5$) nuclear
matter.} \label{extensive}
\end{figure}

Results of the shear viscosity $\eta$ and specific shear viscosity
$\eta/s$, where $s$ is the entropy density, are shown in
Fig.~\ref{extensive}. The temperature dependence of the shear
viscosity is similar to that in Ref.~\cite{Shi03} at different
densities, while $\eta$ increases with increasing density especially
at lower temperatures due to the strong Pauli blocking effect. The
specific shear viscosity decreases with increasing temperature, and
it is similar in both magnitude and trend to those obtained from BUU
calculations using the Green-Kubo formula~\cite{Li11}. It is
interesting to see that at higher temperatures the specific shear
viscosity is about $4\sim5$ times the lower limit from Ads/CFT
calculation~\cite{Kov05}, which is already close to that of QGP
extracted from the study using a viscous hydrodynamical
model~\cite{Son11}. At lower temperatures the specific viscosity
increases with increasing density due to the Pauli blocking effect,
while at higher temperatures the dependence on the density is rather
weak.

\begin{figure}[t]
\includegraphics[scale=0.8]{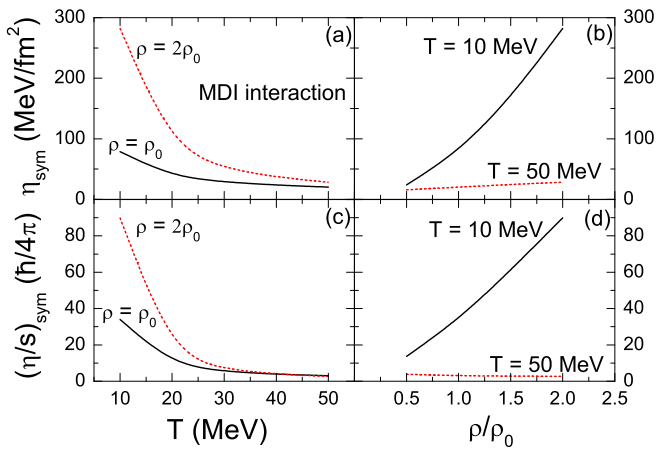}
\caption{(color online) Density and temperature dependence of
symmetry shear viscosity ((a) and (b)) and symmetry specific shear
viscosity ((c) and (d)).} \label{sym}
\end{figure}

Due to the sharper momentum distribution of neutrons compared to
that of protons in neutron-rich nuclear matter, the total shear
viscosity is dominated by neutrons which have a longer relaxation
time in asymmetric nuclear matter compared to that in symmetric
nuclear matter. This is confirmed in Fig.~\ref{extensive} that both
the shear viscosity and the specific shear viscosity are larger in
neutron-rich nuclear matter. In addition, it was seen~\cite{Xu11}
that both the shear viscosity and specific shear viscosity satisfy
the parabolic approximation with respect to the isospin asymmetry,
i.e.,
\begin{eqnarray}
\eta(\rho,T,\delta) &\approx& \eta(\rho,T,\delta=0) +
\eta_{sym}(\rho_,T) \delta^2,\\
\left(\frac{\eta}{s}\right)(\rho,T,\delta) &\approx&
\left(\frac{\eta}{s}\right)(\rho,T,\delta=0) +
\left(\frac{\eta}{s}\right)_{sym}(\rho_,T) \delta^2.
\end{eqnarray}
Similar to the symmetry energy, the second-order coefficient can
thus be defined as the symmetry shear viscosity or the symmetry
specific shear viscosity. The density and temperature dependence of
them are shown in Fig.~\ref{sym}. It is seen that both the symmetry
shear viscosity and symmetry specific shear viscosity decrease with
increasing temperature. At lower temperatures, both of them increase
with increasing density. At higher temperatures, the density
dependence is rather weak. $\eta_{sym}$ and
$\left(\frac{\eta}{s}\right)_{sym}$ are important quantities in
understanding transport properties of neutron-rich nuclear matter,
and they deserve further studies in the future.

\section{Summary and outlook}
\label{summary}

Using a relaxation time approach, I studied the shear viscosity and
specific shear viscosity of hot neutron-rich nuclear matter as that
formed in intermediate-energy heavy-ion collisions by using an
isospin- and momentum-dependent interaction. It is found that the
specific shear viscosity decreases with increasing temperature, and
it increases with increasing density at lower temperatures due to
the strong Pauli blocking effect. Furthermore, both the shear
viscosity and specific shear viscosity are found to increase with
increasing isospin asymmetry of nuclear matter and roughly satisfy
the parabolic approximation. The second-order coefficient in the
expansion of the isospin asymmetry, which is defined as the symmetry
shear viscosity or the symmetry specific shear viscosity, has also
been studied.

\end{document}